\documentclass[twocolumn]{emulateapj}
\usepackage{amsmath}
\usepackage{amssymb}
\usepackage{hyperref}
\usepackage{comment}
\usepackage{color}
\usepackage{natbib}
\usepackage{verbatim}
\usepackage{graphicx}
\usepackage{soul}
\usepackage{rotating}
\usepackage{lineno}
\usepackage{xspace}

\usepackage{aas_macros}

\newcommand{\gammaRays}{{$\gamma$ rays}\xspace}
\newcommand{\gammaRayHyph}{{$\gamma$-ray}\xspace}

\shorttitle{Millisecond Pulsars in Dwarf Spheroidal Galaxies }
\shortauthors{M. Winter, G. Zaharijas, K. Bechtol, \& J. Vandenbroucke}

\begin{document}

\title{Estimating the GeV Emission of Millisecond Pulsars in Dwarf Spheroidal Galaxies}

\author{
Miles Winter\altaffilmark{1,*}, 
Gabrijela Zaharijas\altaffilmark{2,$\dagger$}, 
Keith Bechtol\altaffilmark{1}, 
Justin Vandenbroucke\altaffilmark{1}
}
\altaffiltext{*}{Co-first author, \href{mailto:winter6@wisc.edu}{winter6@wisc.edu}}
\altaffiltext{$\dagger$}{Co-first author, \href{gabrijela.zaharijas@ung.si}{gabrijela.zaharijas@ung.si}}
\altaffiltext{1}{Wisconsin  IceCube  Particle  Astrophysics  Center, University  of  Wisconsin,  Madison,  WI  53706,  USA}
\altaffiltext{2}{Istituto Nazionale di Fisica Nucleare - Sezione Trieste, Padriciano 99, I-34149 Trieste, Italy}

\begin{abstract}

We estimate the conventional astrophysical emission from dwarf spheroidal satellite galaxies (dSphs) of the Milky Way, focusing on millisecond pulsars (MSPs), and evaluate the potential for confusion with dark matter (DM) annihilation signatures at GeV energies.
In low-density stellar environments, such as dSphs, the abundance of MSPs is expected to be proportional to stellar mass.
Accordingly, we construct the \gammaRayHyph luminosity function of MSPs in the Milky Way disk, where $>90$ individual MSPs have been detected with the \textit{Fermi} Large Area Telescope (LAT), and scale this luminosity function to the stellar masses of 30 dSphs to estimate the cumulative emission from their MSP populations.
We predict that MSPs within the highest stellar mass dSphs, Fornax and Sculptor, produce a \gammaRayHyph flux $>500$~MeV of $\sim10^{-11}$~ph~cm$^{-2}$~s$^{-1}$, which is a factor $\sim10$ below the current LAT sensitivity at high Galactic latitudes.
The MSP emission in ultra-faint dSphs, including targets with the largest J-factors, is typically several orders of magnitude lower, suggesting that these targets will remain clean targets for indirect DM searches in the foreseeable future.
For a DM particle of mass 25~GeV annihilating to $b$ quarks at the thermal relic cross section (consistent with DM interpretations of the Galactic Center excess), we find that the expected \gammaRayHyph emission due to DM exceeds that of MSPs in all of the target dSphs.
Using the same Milky Way MSP population model, we also estimate the Galactic foreground MSP coincidence probability along the same sightlines to the dSphs.

\end{abstract}


\maketitle

\section{Introduction}  \label{sec:intro}

Searches for the annihilation products of dark matter (DM) are now testing significant portions of the theoretically motivated parameter space for weakly interacting massive particles.
The rapid progress of indirect DM searches can be attributed to a large number of astrophysical probes that have become available over the last decade \citep[reviewed by][]{Gaskins:2016cha}.
Among these, the Large Area Telescope (LAT) on board \textit{Fermi} has played a vital role due to its full-sky coverage, sensitivity, and energy range relevant for DM searches at the electroweak scale. 
LAT data have been used for numerous DM searches involving a variety of astrophysical objects, including dwarf spheroidal satellite galaxies (dSphs) of the Milky Way (MW).

dSphs are especially promising targets for indirect DM searches due to their (1) substantial DM content \citep[e.g.,][]{mat98, Simon:2007dq} and proximity, (2) distribution over a range of Galactic latitudes, including regions with low diffuse foreground emission, and (3) dearth of non-thermal production mechanisms.
No \gammaRayHyph signal has been conclusively associated with dSphs, either individually or as a population, and the corresponding upper limits have been used to set competitive constraints on DM annihilation \citep[summarized by][]{Charles:2016pgz}.
For example, a joint analysis of 15 dSphs with 6 years of LAT data excluded DM particles annihilating at the canonical thermal relic cross section in some annihilation channels for DM masses up to 100 GeV \citep{ack15}. 

Although the non-DM \gammaRayHyph emission from dSphs is expected to be low, no empirical measurement and few quantitative estimates for this contribution have been previously available.
dSphs have old stellar populations \citep[e.g.,][]{2012ApJ...753L..21B,2014ApJ...789..147W} and low gas content \citep{2009ApJ...696..385G,2014ApJ...795L...5S}, and therefore contain few sites for non-thermal radiation from cosmic-ray (CR) interactions.
However, their ancient stellar populations might include small populations of \gammaRayHyph-emitting millisecond pulsars (MSPs), which have characteristic ages of several Gyr based upon their measured spin periods and period derivatives ($\tau \equiv P / 2 \dot{P}$).
MSPs are luminous sources that account for nearly half of LAT-detected pulsars.
In addition, 25 MW globular clusters, which have similar-age stellar populations to dSphs, have been detected by the collective emission of their MSP populations \citep{abd10,Hooper:2016rap}.

MSPs exhibit hard spectral indices $\sim1.5$ and spectral cut-offs around 3~GeV \citep{abd13,cho14}. 
As a consequence, their intensity peaks in the GeV range, where the LAT sensitivity is highest.
The characteristic spectral shape of MSPs is also similar to that of the Galactic Center excess, and many authors have investigated the contribution of MSPs to that signal \citep[e.g.,][]{Abazajian:2012pn,Brandt:2015,bar15,Lee:2015fea,hoo15}.
In this Letter, we estimate the conventional astrophysical emission intrinsic to dSphs, focusing on MSPs, and evaluate the potential for confusion with DM annihilation signatures at GeV energies.

\section{MSP Formation Mechanisms and the Adopted Strategy}\label{sec:formation}

MSPs are neutron stars that have been spun up to millisecond spin periods via mass accretion from a binary companion \citep{alp82}. 
Different MSP formation mechanisms are thought to dominate in various stellar environments.
The classic ``primordial'' channel begins with a close stellar binary born from a single gas cloud with an extreme mass ratio between the two stars.
These binary systems remain bound after the supernova explosion of the more massive star. 
In a later phase, during mass transfer from the companion star to the neutron star, the binary becomes visible as a low-mass X-ray binary (LMXB), and eventually as a MSP accompanied by a low-mass white dwarf \citep{Zwart:2011py}.
For dense systems with high stellar encounter rates, the MSP formation rate is increased by an additional mechanism: the gravitational capture of neutron stars into binary systems \citep{1998MNRAS.301...15D}.
This ``dynamical'' channel is expected to dominate in globular clusters \citep{Hui:2010vt}.

In galaxy disks, the number of LMXBs scales linearly with stellar mass of the host galaxy \citep{gil04}, suggesting that primordial formation is dominant.
dSphs also have low-density stellar environments, and therefore, it is likely that most LMXBs in dSphs have a primordial origin.
While any direct comparison between the number of LMXBs and MSPs is uncertain, due to the different lifetimes of these two evolutionary stages, it is reasonable to expect the population of field MSPs to scale with stellar mass of the host galaxy if it does so for the progenitor systems.
Thus far, \citet{2005MNRAS.364L..61M} detected five LMXBs in a deep \textit{Chandra} survey of the Sculptor dSph, implying that MSPs may also be present.
No pulsars have been found yet in Ursa Minor, Draco, or Leo~I with searches at 350~MHz using the Green Bank Radio Telescope \citep{Rubio-Herrera:2013}.

Our strategy to predict the \gammaRayHyph emission of MSPs in dSphs will be the following.
First, we construct the \gammaRayHyph luminosity function (LF) of MSPs in the MW disk (Section~\ref{sec:calculation}).
Second, we scale this LF from the stellar mass of the MW to the stellar masses of dSphs and compute the cumulative emission of the MSP population (Section~\ref{sec:dsphs}).
This approach should yield upper limits on the MSP emission because it implicitly assumes that MSPs formed within dSphs do not escape, and that the ages of MSPs in the MW disk and dSphs are similar.

\subsection{Neutron Star Escape}

Neutron stars receive a ``kick'' in the supernova events in which they originate. 
Predicted kick velocity distributions vary widely in the literature: \citet{Hooper:2013nhl} and \citet[][for long lived MSPs]{Cordes:1997my} find 10--50 km s$^{-1}$, \citet{hob05} and \citet{tos99} find $\sim85\pm13$ km s$^{-1}$, while \citet{lyn98} find a higher velocity range $\sim130\pm30$ km s$^{-1}$.
If unbound, MSPs would leave these systems in $t_{\rm esc} \sim 10^8 (R/{\rm 1~kpc}) (10~{\rm km s}^{-1} /v_{\rm kick})$ yr, i.e., much shorter than the typical stellar ages of dSphs. 
The kick-velocity estimates above are larger than the typical stellar velocity dispersion of dSphs, $\sim 10$ km s$^{-1}$ in dSphs \citep{mcc12}, but are comparable to the escape velocity. 
In particular, the enclosed masses of classical dwarfs within 600 pc from the center are (2--7)$ \times 10^7$ M$_{\odot}$ \citep{Walker:2007ju}, corresponding to an escape velocity $v_{\rm esc}=\sqrt{2GM_{600}/r_{600}} \sim 30$ km s$^{-1} \sim v_{\rm kick}$. 
This suggests that a fraction of MSPs should be retained within the virial radii of dSphs, which are assumed to be at kpc distances with a contained mass of $\sim10^{8-9}$ M$_{\odot}$ \citep[see also the calculations of ][]{Dehnen:2005fs}.\footnote{MSPs with larger kick velocities would attain higher-eccentricity orbits and this may provide a means to constrain the mass profile of dSphs beyond the radii of their normal stellar populations.}

An MSP orbiting at 600~pc from the center of a dSph located 80~kpc away would span an apparent $0\fdg4$ angular size, which is within the 95\% confinement radius of LAT \texttt{Pass8 FRONT} events at 2~GeV.\footnote{\url{https://www.slac.stanford.edu/exp/glast/groups/canda/lat\_Performance.htm}} 
Depending on the distance to a given dSph and the photon event class used for the analysis, a slight signal extension might be present.

\subsection{Stellar Population Comparison}

Star formation in most MW dSphs ended several Gyr ago \citep[e.g.,][]{2012ApJ...753L..21B,2014ApJ...789..147W}, whereas the MW disk remains active.
The formation of LMXBs, and accordingly MSPs, is thought to peak most strongly in the first $0.5$~Gyr after star formation \citep{fra08}, suggesting that MSPs in dSphs are potentially older and less luminous \citep{Hooper:2016rap} than the average Galactic MSP.

However, other factors might increase the MSP population in dSphs.
Neutron star production could be enhanced in low metallicity systems relative to the Galactic field by $\sim20$\% \citep{Ivanova:2007bu}.
Also, two dSphs host their own globular clusters: five in Fornax \citep{2016A&A...590A..35D} and one in Eridanus~II \citep{crn16}.
The abundance of MSPs might be higher in these systems due to more frequent stellar encounters.

\section{Derivation of a Stellar-Mass-normalized MSP Luminosity Function}\label{sec:calculation}

\subsection{MSP Sample}\label{sec:sample}

\begin{figure*}
\includegraphics[width=.5\textwidth]{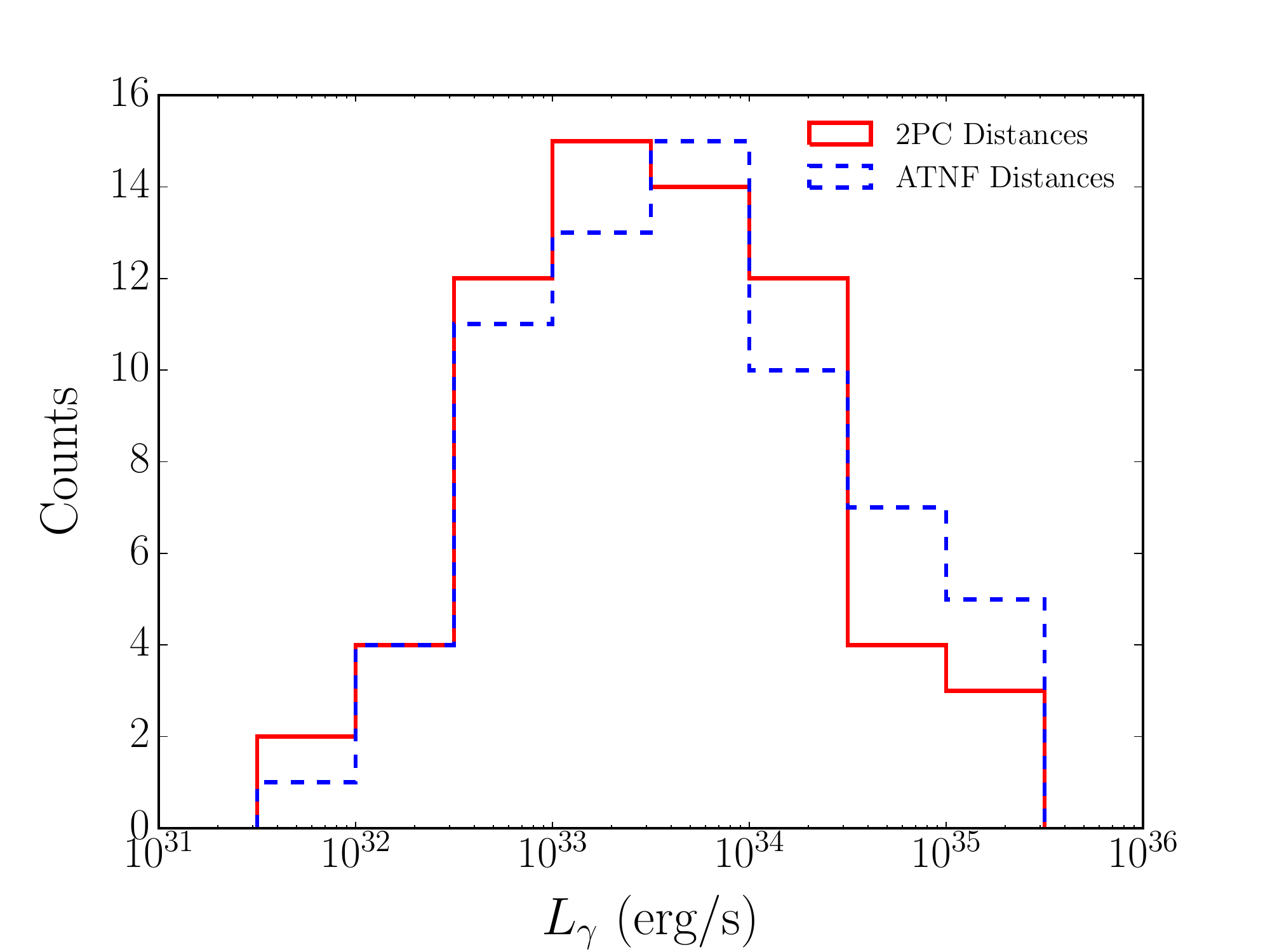}
\includegraphics[width=.5\textwidth]{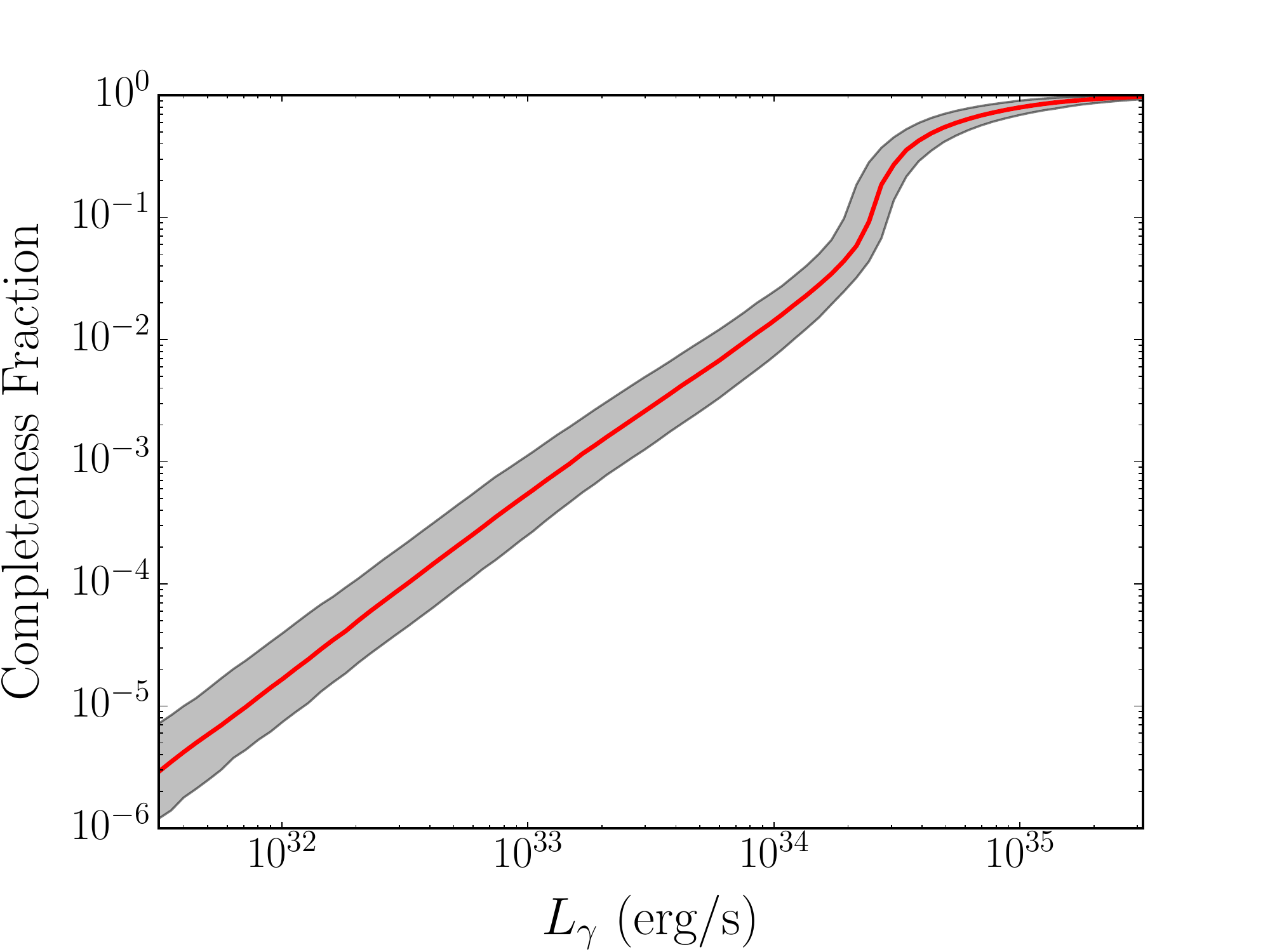}
\caption{\textit{Left}: Luminosity (0.1--100~GeV) distribution of LAT detected MSPs calculated for 2PC and ATNF reported distances. \textit{Right}: Estimated LAT survey completeness fraction as a function of \gammaRayHyph luminosity (0.1--100~GeV). The red curve and gray band indicate the median and inner 68\% interval of the completeness fraction across the MC realizations. \label{fig:complete}}
\end{figure*}

Our pulsar sample consists primarily of MSPs in the Second LAT Pulsar Catalog \citep[2PC;][]{abd13}, although all LAT-detected MSPs were considered.\footnote{\url{https://confluence.slac.stanford.edu/display/GLAMCOG/Public+List+of+LAT-Detected+Gamma-Ray+Pulsars}\label{fn:lat_msps}}
Pulsars residing within globular clusters, as well as those lacking distance measurements, were both excluded, leaving 66 MSPs in our sample. 

The luminosities of Galactic MSPs often have large uncertainties because accurate distance measurement is difficult.
Distance measurements based on dispersion measure suggest that for roughly 75\% of the directions in the sky the accuracy is no better than a factor of 1.5 to 2 \citep{sch12}. 
We compare luminosities calculated with distances from 2PC and from the Australian Telescope National Facility Pulsar Catalog \citep[ATNF;][]{man05}\footnote{\url{http://www.atnf.csiro.au/people/pulsar/psrcat/}} in Figure \ref{fig:complete}.

\subsection{Incompleteness Correction}\label{sec:completeness}

As a first step towards constructing the Galactic MSP LF we perform a Monte Carlo (MC) incompleteness correction and apply it to our MSP sample. 
In our model, we assume an exponential spatial distribution for MSPs in the MW disk:
\begin{equation}
\rho(R,z)\propto e^{(-R/R_0)}e^{(-|z|/z_0)},
\label{eqn:mw_dist}
\end{equation}

\noindent where $R$ is the radial distance from the Galactic center and $z$ is the vertical scale height above the Galactic plane.
While other authors model the radial distribution of Galactic MSPs using a Gaussian density profile \citep{fau10}, \citet{gre13} find that their results are fairly insensitive to the selected radial law. 
We make no attempt to model a special population of MSPs in the Galactic bulge; our population of interest is field MSPs.

To account for systematic uncertainties both in the spatial distribution of Galactic MSPs, as well as the effective selection function of the LAT pulsar survey, we repeat the following MC procedure many times with different input parameter sets.
First, following the MSP population model of \citet{gre13}, we draw radial and vertical scale lengths from log-normal distributions defined by $R_0=3^{+3}_{-1}$~kpc and $z_0=0.6^{+0.6}_{-0.3}$~kpc, respectively (see their Table~3).
Using these scale lengths, we generate $10^7$ MSPs at random locations consistent with the spatial distribution of Equation \ref{eqn:mw_dist} and assuming azimuthal symmetry in Galactocentric coordinates.
We then determine the apparent Galactic coordinates $(l, b)$ of each MSP as viewed from the Sun's position at $(R,\phi,z) = (8.5\text{ kpc},0\text{ rad}, 20\text{ pc})$.

Each MSP $(l,b)$ coordinate is mapped to an effective flux detection threshold using the sensitivity curve in Figure~17 of 2PC \citep{abd13}, which is expressed as a function of Galactic latitude.
This direction-dependent flux threshold partially accounts for variations in the intensity of diffuse Galactic emission.
We model systematic uncertainty in the 2PC selection function by drawing a direction-dependent flux threshold curve between the 10\% and 90\% percentile sensitivity range for each realization of the MSP population (sampled from a uniform distribution).
The survey completeness is evaluated as the number of detectable MSPs at a given luminosity divided by the total number of simulated MSPs.

After calculating the detection efficiency for $10^5$ sets of spatial parameters and flux detection thresholds, we arrive at the completeness function shown in Figure \ref{fig:complete}.

\subsection{\gammaRayHyph Luminosity Function}\label{sec:lumfunction}

\begin{figure}
\centering
\includegraphics[width=.5\textwidth]{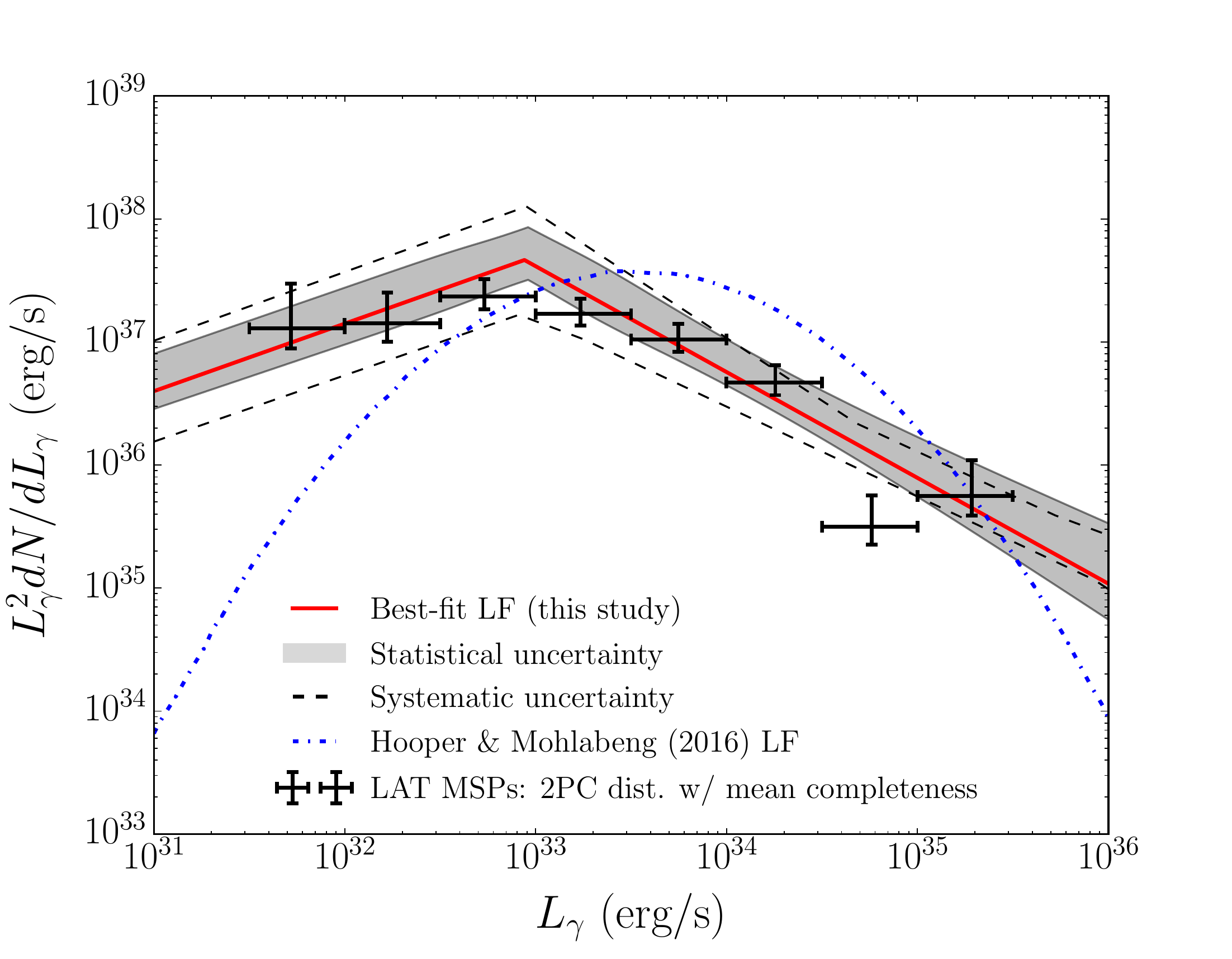}
\caption{MSP \gammaRayHyph luminosity function (0.1--100~GeV) normalized to the stellar mass of the MW.
Data points include a incompleteness correction applied to LAT MSP sample assuming 2PC distances. 
Error bars correspond to statistical uncertainty associated with the finite number of LAT-detected MSPs. 
The shaded gray band represents the $1\sigma$ statistical uncertainty on the broken power-law fit to these data and dashed gray lines represent the systematic uncertainty envelope (distances to LAT-detected MSPs, spatial distribution of Galactic MSPs, and effective selection function of LAT pulsar catalog). The blue dashed curve represents the best-fit LF of \citet{hoo15} normalized to the cumulative luminosity of our best-fit LF.
\label{fig:l2dndl}}
\end{figure}

We derive the MSP \gammaRayHyph LF using a Markov Chain Monte Carlo (MCMC) assuming a broken power law parametrization.
During the fit, the median incompleteness correction (Section~\ref{sec:completeness}) is applied to the MSPs in each luminosity bin of our sample.
After determining the LF posterior, we normalize to unit stellar mass, assuming a MW stellar mass of $7\times10^{10}M_\sun$ \citep{mal96}. 
The resultant LF is shown in Figure \ref{fig:l2dndl}. 

To estimate the systematic uncertainty, we re-fit the LF pairing luminosities calculated from 2PC and ATNF distances with the upper and lower limits of the incompleteness correction. 
These four bracketing cases establish the upper and lower systematic envelope on the LF. 
At high luminosity, uncertainty on the LF is driven by the small number of extremely luminous Galactic MSPs, since the LAT census of such sources is expected to be largely complete. 
Systematic uncertainty dominates at low luminosity, due in part to the large and imperfectly known incompleteness correction that must be applied.

Our model predicts that the main contribution comes from MSPs with luminosities $\sim10^{33}~\text{erg}~\text{s}^{-1}$. 
Both our incompleteness correction and LF are in reasonable agreement with \citet{hoo15}; their best-fit LF for field MSPs peaks around $3 \times 10^{33}~\text{erg}~\text{s}^{-1}$ in $L_{\gamma}^2 dN/dL_{\gamma}$.

\section{Cumulative MSP Emission Towards dSphs}\label{sec:dsphs}

Our target sample, summarized in Table~\ref{tbl:dsph_flux}, includes both spectroscopically confirmed dSphs and recently reported dSph candidates, which we collectively refer to as ``dSphs" for simplicity.
Since our model requires the stellar mass of the host galaxy as an input, we include only targets with published stellar masses.
We first discuss the emission from MSPs within dSphs, followed by the MW foreground contribution.

\subsection{Internal MSP Emission of dSphs}

\begin{figure*}
\centering
\includegraphics[width=1.\textwidth]{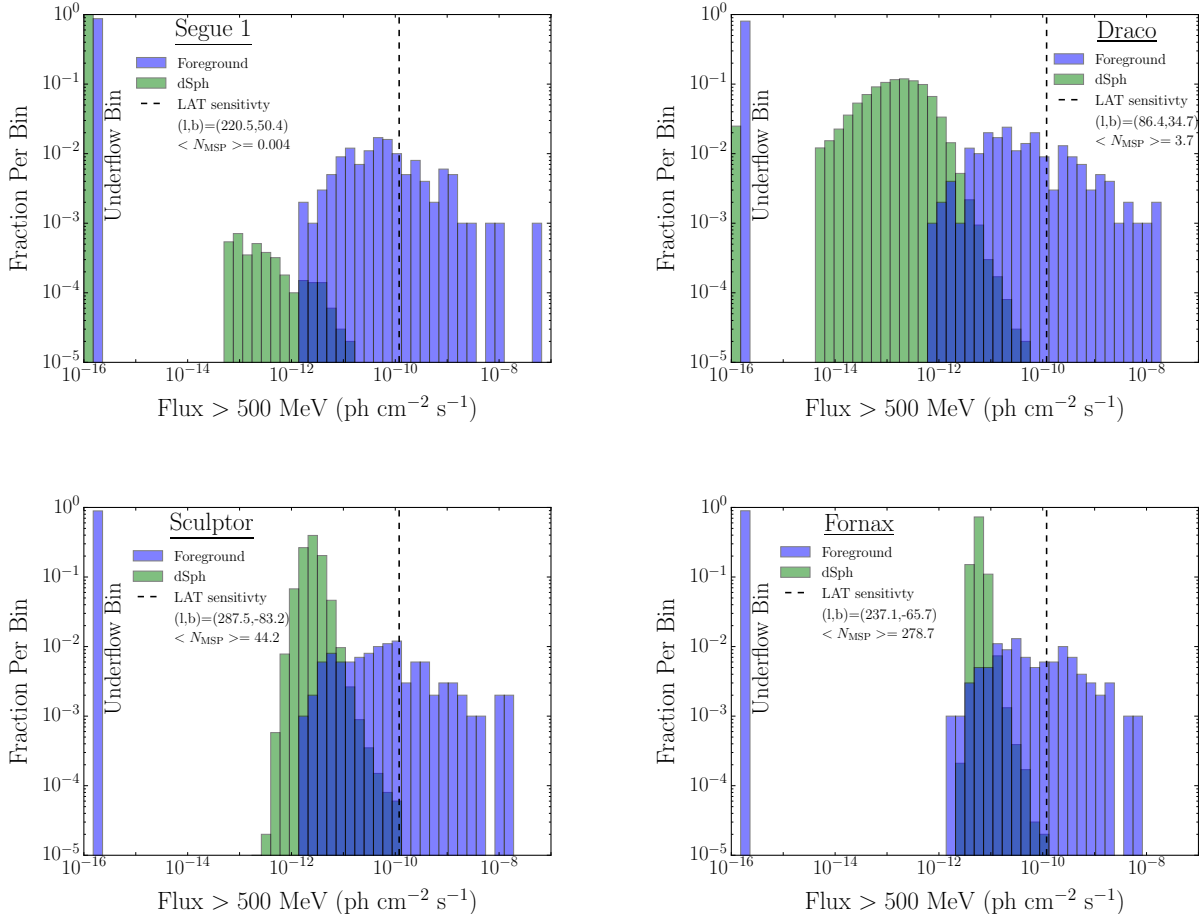}
\caption{Predicted flux distributions illustrating the effect of Poisson fluctuations (i.e., ``shot noise'') in the number of MSPs towards four representative dSphs. 
Green histograms represent the contribution from MSPs within each dSph, and blue histograms represent the contribution from MW foreground MSPs along the same line of sight (integrated within a $1\fdg$ radius of the dSph location).
Bins labeled ``underflow'' represent the fraction of realizations without a MSP flux contribution (no MSPs with $L_{0.1-100~\rm{GeV}}>10^{31}~\text{erg}~\text{s}^{-1}$). 
For comparison, the dashed black vertical line indicates the typical LAT sensitivity at high Galactic latitudes ($2\sigma$ upper limit).
\label{fig:pois}}
\end{figure*}

\begin{figure*}
\centering
\includegraphics[width=1.\textwidth]{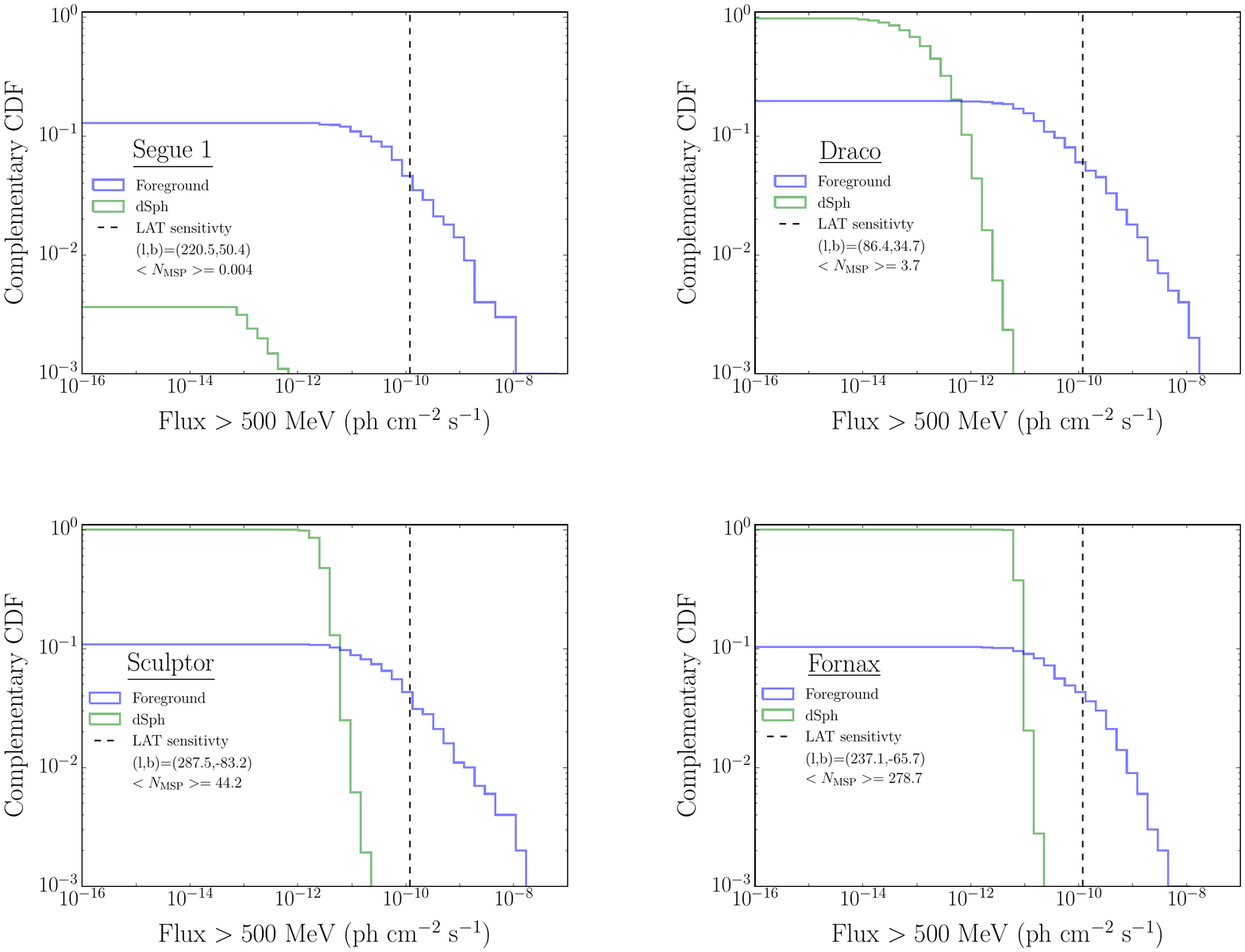}
\caption{Same as Figure~\ref{fig:pois}, but representing predicted MSP flux distributions with complementary cumulative distribution functions. 
\label{fig:survival}}
\end{figure*}

For each of the 30 dSphs in our sample, we scale the LF of Galactic MSPs (Section~\ref{sec:lumfunction}) to the dSph stellar mass.
The expectation value for the cumulative MSP luminosity is found by integrating the LF, $dN/dL_\gamma$:

\begin{equation}
\langle L_{\gamma,{\rm MSP}} \rangle =  \frac{M_{*, {\rm dSph}}}{M_{*, {\rm MW}}} \int_{L_{\rm min}}^{L_{\rm max}} L_\gamma \frac{dN}{dL_\gamma} dL_\gamma,
\end{equation}
where $L_{\rm min}=10^{31}~\text{erg}~\text{s}^{-1}$ and $L_{\rm max}=10^{36}~\text{erg}~\text{s}^{-1}$.

dSphs have small enough stellar populations that the number of luminous MSPs in a particular dSph may exhibit substantial ``shot noise''.
To model this effect, we evaluate the cumulative MSP flux including Poisson fluctuations in the number of MSPs within multiple luminosity intervals.
We repeat this process $10^4$ times for each dSph to construct a PDF of the expected flux.
As shown in Figures~\ref{fig:pois} and \ref{fig:survival}, the spread in cumulative MSP flux for a low stellar mass dSph, such as Segue 1, spans several order of magnitude, whereas high stellar mass dSphs, such as Fornax and Sculptor, have sharply peaked flux PDFs. 
The cumulative MSP flux in an ultra-faint dSph is likely dominated by a handful of luminous sources, but the relative impact of such statistical variations diminishes with increasing stellar mass.

\subsection{Galactic Foreground MSP Emission}

Using the same MW MSP population model discussed in Section~\ref{sec:sample}, we also estimate the probability of confusion due to the chance alignment of foreground MSPs along the same sightlines to the target dSphs.
We begin by randomly drawing luminosities from the best-fit LF (normalized to the stellar mass of the MW). 
Next, the luminosity values are assigned to random MSP locations within the MW according to the fiducial spatial model described in Section~\ref{sec:completeness}. 
We then compute the cumulative flux within a $1\fdg0$ radius (comparable to the LAT PSF for 1~GeV \gammaRays) around the location of each dwarf in our sample. 
The resulting foreground MSP flux distributions for four representative dSphs are shown in Figures~\ref{fig:pois} and \ref{fig:survival}.

\subsection{Results and Discussion}\label{sec:results}

For each of the 30 dSphs in our sample, we compute the expectation value and uncertainty on the cumulative MSP flux by adding in quadrature
\begin{enumerate}
\item statistical uncertainty due to the finite number of LAT-detected Galactic MSPs, 
\item systematic uncertainty in the luminosities of Galactic MSPs due to distance uncertainties, 
\item systematic uncertainty in the detection efficiency of Galactic MSPs due to their unknown spatial distribution and the effective flux threshold of the LAT pulsar catalog, and
\item uncertainty in the stellar mass of the host dSph. 
\end{enumerate}

\noindent We then compute flux upper limits taking into account Poisson fluctuations in the number of luminous MSPs (using the best-fit LF). While stellar mass uncertainty is dominant for select dSphs, we find that systematic uncertainty, together with Poisson fluctuations, have the greatest impact on the predicted signals from both ultra-faint and classical dSPhs. The results of these calculations, summarized in Table~\ref{tbl:dsph_flux}, suggest that high stellar mass dSphs are likely to host modest MSP populations.   
Consistent with our initial assumptions, we find the mean number of predicted MSPs to be higher than a few in all dSphs where LMXBs or LMXB candidates have been detected.
However, even for the largest classical dSph, Fornax, the predicted MSP flux $> 500$~MeV is $6.0^{+9.7}_{-4.4}\times10^{-12}$~ph~cm$^{-2}$~s$^{-1}$, which is about an order of magnitude below the typical flux upper limits obtained at high Galactic latitudes after six years of the LAT survey, $\sim10^{-10}$ ph cm$^{-2}$ s$^{-1}$ \citep{ack15}.
Perhaps more importantly, Draco and Ursa Minor, two classical dSphs with large and well constrained J-factors \citep[e.g.,][]{Geringer-Sameth:2014yza}, have estimated MSP fluxes another order of magnitude fainter than Fornax.
The MSP emission of ultra-faint dSphs, such as Segue 1, is typically below $10^{-13}$~ph~cm$^{-2}$~s$^{-1}$ and likely beyond the reach of \gammaRayHyph telescopes in the foreseeable future.

\begin{figure*}
\centering
\includegraphics[width=1.\textwidth]{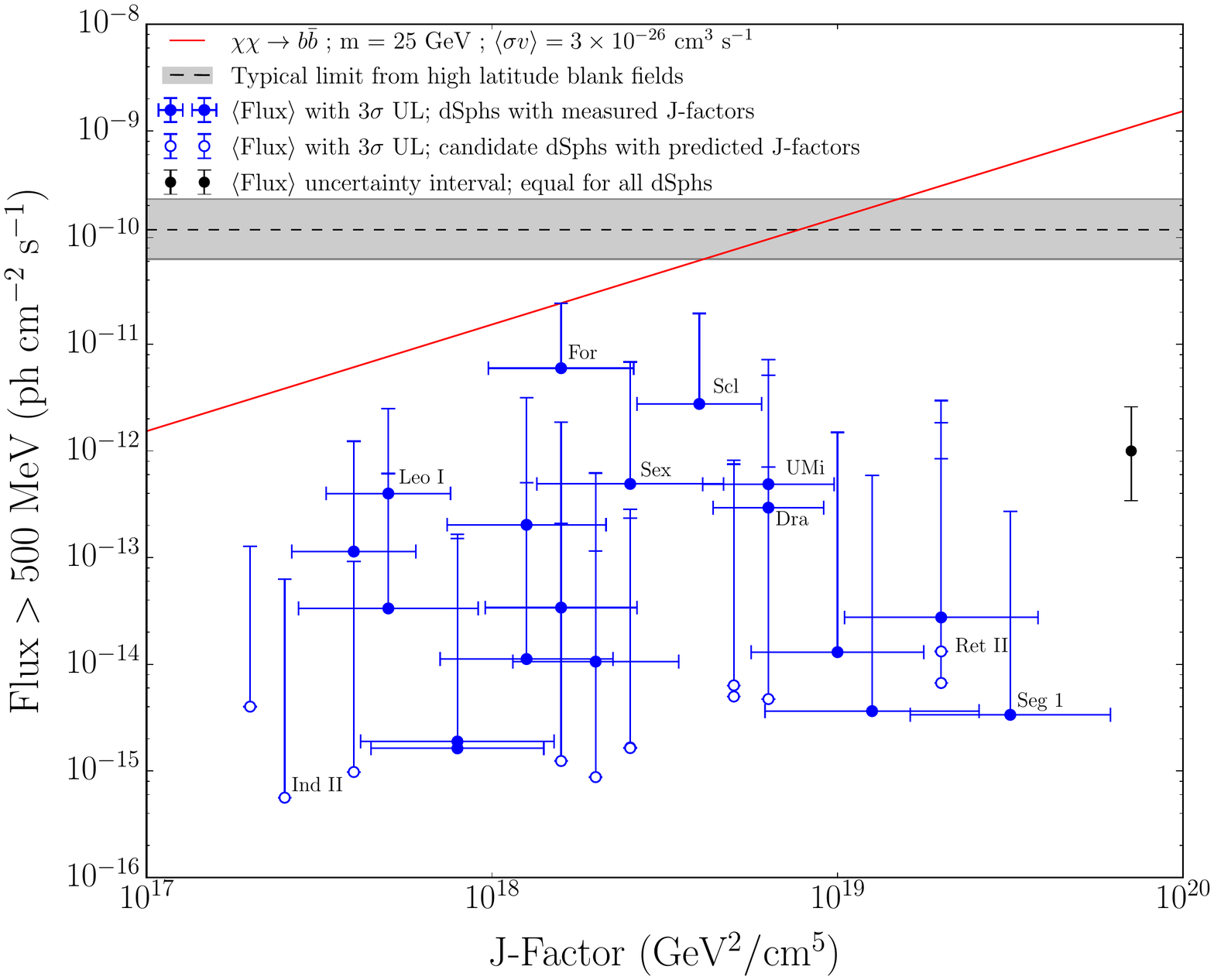}
\caption{Expected \gammaRayHyph flux versus J-factor. 
Blue points indicate expectation values for the predicted MSP emission in 30 MW dSphs and dSph candidates. 
Blue vertical error bars show 3$\sigma$ upper limits on the predicted flux due to Poisson fluctuations alone. 
Additional uncertainty is attributed to uncertainty on the LF of MW MSPs (see Section~\ref{sec:results}).
This contribution spans the same logarithmic interval for all targets due to the assumed linear scaling with stellar mass, and is represented by the rightmost set of black error bars.
J-factor uncertainties are shown for kinematically confirmed dSphs only. The predicted strengths of the MSP and DM annihilation (red line) signals have different dependence on the J-factor. This distinction offers another means to distinguish a putative DM signal from conventional astrophysical sources (e.g., using a joint-likelihood analysis). The gray shaded band represents the typical $2\sigma$ upper limit derived in high Galactic latitude blank fields after 6 years of the LAT survey. The red curve represents a DM annihilation model that is consistent with both DM interpretations of the Galactic Center Excess and the characteristic spectral shape of MSPs. \label{fig:sum_fig}}
\end{figure*}

To more easily compare \gammaRayHyph signals from MSPs and DM annihilation, in Figure~\ref{fig:sum_fig} we show the expected MSP flux versus J-factor for the host dSph.
The potential for confusion is greatest between models with similar \gammaRayHyph spectra, so we consider a DM model that produces a \gammaRayHyph energy flux peaking at $\sim1$~GeV, namely $\chi\chi \rightarrow b \bar{b}$ with a DM particle mass of 25 GeV. 
This DM model is also compatible with DM interpretations of the Galactic Center excess \citep[e.g.,][and references therein]{Calore:2014xka}.
In addition to being below the current LAT sensitivity threshold, we predict that MSP emission in each of the dSphs is also below the annihilation signal expected from this DM model for an annihilation cross section at the thermal relic value $\langle \sigma v \rangle = 3 \times 10^{-26} {\rm cm}^3 {\rm s}^{-1}$.
 
Figure~\ref{fig:sum_fig} also demonstrates that the GeV emission from MSPs and from DM annihilation follow different scaling relations across the dSph population.
The targets with the largest J-factors are mainly nearby ultra-faint dSphs with $M_{*} < 10^{4} M_{\odot}$, whereas we argue that dSphs with $M_{*} \sim 10^{7} M_{\odot}$ are most likely to host detectable \gammaRayHyph MSP populations.
This distinction offers another means to distinguish a putative DM signal from conventional astrophysical sources (e.g., using a joint-likelihood analysis).

We find that Galactic foreground MSP emission is potentially a non-negligible source of confusion for a small fraction of dSphs, even at the current LAT sensitivity.
However, for a typical target, there is no foreground MSP with luminosity exceeding $L_{0.1-100~\rm{GeV}}=10^{31}~\text{erg}~\text{s}^{-1}$ in $\approx90\%$ of model realizations.
For dSphs with $M_{*} > 10^{4} M_{\odot}$, the contribution from MSPs within the dSph is predicted to exceed the foreground in most cases (e.g., Fornax, Sculptor, Draco).
Otherwise, the MSP emission internal the dSph is likely to be sub-dominant (e.g., Segue~1).
Several LAT analyses have already employed empirical background modeling methods to partially account for sub-threshold point sources coincident with dSphs \citep[e.g.,][]{2011PhRvL.107x1303G,Ackermann:2013yva,Carlson:2014nra}.

\section{Cosmic-ray Induced Emission in dSphs}

The non-thermal emission in dSphs related to CR interactions is highly uncertain due to our present lack of knowledge regarding the acceleration and confinement of CRs in such environments.
Still, it is possible to obtain an estimate using the quasi-linear scaling relation between \gammaRayHyph luminosity and tracers of the star formation rate \citep[SFR;][]{Ackermann:2012vca}.
Considering the most massive dSph, Fornax \citep[${\rm SFR} \sim 10^{-4}$~M$_{\odot}$~yr$^{-1}$;][]{2012A&A...544A..73D}, and using the SMC as the nearest point of comparison with $L_{0.1-100~{\rm GeV}} \sim 1 \times 10^{37}$~erg~s$^{-1}$ and ${\rm SFR} \sim 10^{-1}$~M$_{\odot}$~yr$^{-1}$ \citep{2004AJ....127.1531H}, the predicted CR-induced luminosity of Fornax is $L_{0.1-100~{\rm GeV}} \sim 1 \times 10^{34}$~erg~s$^{-1}$. 
This estimate is comparable to the expected MSP luminosity, but should be regarded as an upper limit considering the low gas content of dSphs \citep[${\rm M}_{\rm HI} / {\rm M}_{*} \lesssim 10^{-2}$;][]{2014ApJ...795L...5S} compared to the SMC \citep[${\rm M}_{\rm HI} / {\rm M}_{*} \sim 1$;][]{mcc12}.

In addition, the interval between CR acceleration events in dSphs might be comparable to the CR confinement time, in which case the CR population may not reach an equilibrium state.
Following \citet{2015ApJ...805L...2C}, the rates of core-collapse and type Ia supernovae for a galaxy with a constant SFR are $\approx 10^{-2}$~yr$^{-1}$ and $\approx 10^{-13}$~yr$^{-1}$~M$_{\odot}^{-1}$, respectively. 
Applying these rates to Fornax \citep[$M_{*} \approx 2 \times 10^{7}$~M$_{\odot}$;][]{mcc12}, the expected supernova rate is $\sim1$~Myr$^{-1}$.
By comparison, the CR confinement time in the (much larger) MW at GeV energies is $\sim10^{7}$~yr.
\citet{Chen:2015zcw} proposed that afterglows from tidal disruption events might also enhance \gammaRayHyph emission in dSphs, although the duty cycle of such events is likely to be low.

\section{Conclusions} \label{sec:summary}

dSphs have commonly been regarded as astrophysically ``clean'' targets for which the detection of excess \gammaRays would constitute compelling evidence for particle DM.
However, conventional astrophysical emission must be present at some level, and in this Letter we predict the contribution from MSPs. 
Under the assumption that MSPs in both the MW disk and in dSphs originate mainly from primordial binary systems (in contrast with globular clusters), we scale the LF of Galactic MSPs to the stellar masses of dSphs to quantify their MSP populations.

We estimate that MSP emission within the highest stellar mass dSphs, Fornax and Sculptor, is a factor $\sim10$ below the current LAT sensitivity threshold (Figure~\ref{fig:sum_fig}).
The MSP emission within ultra-faint dSphs (including targets with the largest J-factors) is several orders of magnitude lower.
Moreover, for a DM particle of mass 25~GeV annihilating to $b$ quarks at the thermal relic cross section (consistent with DM interpretations of the Galactic Center excess), the expected \gammaRayHyph emission due to DM exceeds that of the MSP population in all of the dSphs considered here.
At the current LAT sensitivity, the more likely source of confusion is a Galactic foreground MSP along the same line of sight, although the probability of this alignment is typically $\lesssim10\%$ per target dSph (Figure~\ref{fig:survival}).

The LAT sensitivity to DM annihilation in dSphs is anticipated to improve by nearly an order of magnitude over the coming decade due to increasing LAT exposure, more precise J-factor measurements from deep spectroscopy, and additional dSph targets discovered in optical surveys such as LSST \citep{Charles:2016pgz}.
These forecasts are based on a combined likelihood analysis weighting dSphs by their J-factors, as appropriate for DM searches. 
Since we do not expect that many nearby and high stellar mass dSphs remain to be discovered, the sensitivity to the MSP contribution may not improve as quickly as for DM signals. 

\section{Acknowledgements}
We are especially indebted to Andy Strong for the insights provided by his GALPLOT code runs. We acknowledge helpful discussions with Andrea Albert, Brandon Anderson, John Beacom, Alessandro Bressan, Alessandro Cuoco, Alex Drlica-Wagner, German Arturo Gomez-Vargas, Dan Hooper, Matthew Kerr, Tim Linden, Dimitry Malyshev, Nestor Mirabal, Pasquale Serpico, David Smith, and Piero Ullio. 
We thank the anonymous referee for constructive suggestions.

\begin{table*}
\centering
\caption{Estimated GeV Contribution from MSPs in MW dSphs \label{tbl:dsph_flux}}
\begin{tabular}{l l c c c c c c c c c c}
\tableline\tableline 
Galaxy\tablenote{dSphs marked with a star, $^*$, contain globular clusters. A dagger, $^\dagger$, indicates that LMXB or LMXB candidates have been detected.} & D(kpc) & $(l,b)$ & $\log_{10}\left(\frac{M_*}{M_\odot}\right)$ & \multicolumn{5}{c}{\underline{$\quad{}\quad{}\log_{10}\left(\text{Flux} >500 \text{MeV} \left[\frac{\text{ph}}{\text{cm}^{2} \text{s}^{1}}\right]\right)\tablenote{Column 1: flux expectation value calculated from the best-fit LF. The uncertainty on the expectation value is computed by adding in quadrature the $1\sigma$ statistical ($^{+0.18}_{-0.12}$ dex) and systematic ($^{+0.67}_{-0.26}$ dex) uncertainty on the LF together with stellar mass uncertainty. Columns 2, 3: 2$\sigma$ and $3\sigma$ upper limits on the predicted flux due to Poisson fluctuations in the number of MSPs. Columns 4, 5: 1$\sigma$ and 2$\sigma$ upper limits on the expected flux from MW foreground MSPs ($>10^{31}$~erg~s$^{-1}$) along the line of sight (integrated within a $1\fdg0$ radius of the dSph location). A dash, $-$, denotes zero predicted MSPs with $L_{0.1-100~{\rm GeV}} > 10^{31}$~erg~s$^{-1}$ at the given confidence level.}\quad{}\quad{}$}} &  $\langle\text{N}_{\rm MSP}\rangle$\tablenote{Predicted number of MSPs with $L_{0.1-100~{\rm GeV}} > 10^{31}$~erg~s$^{-1}$ computed from the best-fit LF} & $\log_{10}\left(\text{J} \left[\frac{\text{GeV}^2}{\text{cm}^5}\right]\right)$\tablenote{For dSphs that are kinematically confirmed to be DM dominated, measured J-factors with uncertainties are quoted from \citet{ack15}. For the remaining targets, we use a predicted J-factor following the distance scaling relation proposed by \citet{drl15}.} & Ref. \\[1ex]
& & & & Mean & \multicolumn{2}{c}{Internal} & \multicolumn{2}{c}{Foreground} & & & \\
& & & & & $2\sigma$ & $3\sigma$ & $1\sigma$ & $2\sigma$ & & &\\[.5ex]
\tableline
Segue 1	&$	23	$&$(	220	,	50	)$&$	2.53	_{-	0.20	}^{+	0.38	}$&$	-14.47	_{-	0.35	}^{+	0.79	}$&$	-	$&$	-12.57	$&$	-	$&$	-12.12	$&$	0.004	$&$	19.5	\pm	0.29	$&	1,5	\\
Tucana III	&$	25	$&$(	315	,	-56	)$&$	2.90	_{-	0.05	}^{+	0.05	}$&$	-14.18	_{-	0.29	}^{+	0.69	}$&$	-	$&$	-12.08	$&$	-13.4	$&$	-11.74	$&$	0.009	$&$	19.3			$&	4	\\
Reticulum II	&$	32	$&$(	266	,	-50	)$&$	3.41	_{-	0.03	}^{+	0.30	}$&$	-13.88	_{-	0.29	}^{+	0.69	}$&$	-13.45	$&$	-11.74	$&$	-13.85	$&$	-11.88	$&$	0.03	$&$	19.3			$&	3,6	\\
Ursa Major II	&$	32	$&$(	152	,	37	)$&$	3.73	_{-	0.23	}^{+	0.23	}$&$	-13.56	_{-	0.37	}^{+	0.73	}$&$	-12.86	$&$	-11.53	$&$	-13.76	$&$	-12.00	$&$	0.06	$&$	19.3	\pm	0.28	$&	2,5	\\
Willman I	&$	38	$&$(	159	,	57	)$&$	3.00	_{-	0.22	}^{+	0.39	}$&$	-14.44	_{-	0.36	}^{+	0.79	}$&$	-	$&$	-12.23	$&$	-	$&$	-12.19	$&$	0.01	$&$	19.1	\pm	0.31	$&	1,5	\\
Coma Ber.	&$	44	$&$(	242	,	84	)$&$	3.68	_{-	0.22	}^{+	0.22	}$&$	-13.89	_{-	0.36	}^{+	0.73	}$&$	-13.22	$&$	-11.83	$&$	-	$&$	-12.32	$&$	0.05	$&$	19.0	\pm	0.25	$&	2,5	\\
Tucana IV	&$	48	$&$(	313	,	-55	)$&$	3.34	_{-	0.06	}^{+	0.08	}$&$	-14.30	_{-	0.29	}^{+	0.69	}$&$	-13.89	$&$	-12.13	$&$	-13.34	$&$	-11.83	$&$	0.02	$&$	18.7			$&	4	\\
Grus II	&$	53	$&$(	351	,	-52	)$&$	3.53	_{-	0.05	}^{+	0.04	}$&$	-14.20	_{-	0.29	}^{+	0.69	}$&$	-13.65	$&$	-12.09	$&$	-12.95	$&$	-11.88	$&$	0.04	$&$	18.7			$&	4	\\
Tucana II	&$	58	$&$(	328	,	-52	)$&$	3.48	_{-	0.14	}^{+	1.01	}$&$	-14.33	_{-	0.32	}^{+	1.22	}$&$	-13.81	$&$	-12.16	$&$	-13.08	$&$	-11.81	$&$	0.03	$&$	18.8			$&	3,6	\\
Bootes I	&$	66	$&$(	358	,	70	)$&$	4.45	_{-	0.06	}^{+	0.09	}$&$	-13.47	_{-	0.29	}^{+	0.70	}$&$	-12.46	$&$	-11.74	$&$	-	$&$	-12.06	$&$	0.3	$&$	18.2	\pm	0.22	$&	1,5	\\
Indus I	&$	69	$&$(	347	,	-42	)$&$	2.90	_{-	0.22	}^{+	0.22	}$&$	-15.06	_{-	0.36	}^{+	0.72	}$&$	-	$&$	-12.94	$&$	-12.58	$&$	-11.57	$&$	0.009	$&$	18.3			$&	3,6	\\
Draco\tablenotemark{$\dagger$}	&$	76	$&$(	86	,	35	)$&$	5.51	_{-	0.10	}^{+	0.10	}$&$	-12.53	_{-	0.30	}^{+	0.70	}$&$	-11.87	$&$	-11.32	$&$	-13.61	$&$	-11.86	$&$	3.7	$&$	18.8	\pm	0.16	$&	2,5	\\
Ursa Minor	&$	76	$&$(	105	,	45	)$&$	5.73	_{-	0.20	}^{+	0.20	}$&$	-12.31	_{-	0.35	}^{+	0.72	}$&$	-11.73	$&$	-11.17	$&$	-12.97	$&$	-11.91	$&$	6.1	$&$	18.8	\pm	0.19	$&	2,5	\\
Sculptor\tablenotemark{$\dagger$}	&$	86	$&$(	288	,	-83	)$&$	6.59	_{-	0.21	}^{+	0.21	}$&$	-11.56	_{-	0.36	}^{+	0.72	}$&$	-11.22	$&$	-10.78	$&$	-	$&$	-12.13	$&$	44.2	$&$	18.6	\pm	0.18	$&	2,5	\\
Sextans	&$	86	$&$(	244	,	42	)$&$	5.84	_{-	0.20	}^{+	0.20	}$&$	-12.31	_{-	0.35	}^{+	0.72	}$&$	-11.76	$&$	-11.20	$&$	-13.63	$&$	-11.86	$&$	7.9	$&$	18.4	\pm	0.27	$&	2,5	\\
Horologium I	&$	87	$&$(	271	,	-55	)$&$	3.38	_{-	0.13	}^{+	0.25	}$&$	-14.78	_{-	0.31	}^{+	0.74	}$&$	-14.34	$&$	-12.55	$&$	-14.11	$&$	-11.87	$&$	0.03	$&$	18.4			$&	3,6	\\
Reticulum III	&$	92	$&$(	274	,	-46	)$&$	3.30	_{-	0.15	}^{+	0.13	}$&$	-14.91	_{-	0.32	}^{+	0.70	}$&$	-	$&$	-12.68	$&$	-13.39	$&$	-11.77	$&$	0.02	$&$	18.2			$&	4	\\
Phoenix II	&$	95	$&$(	324	,	-60	)$&$	3.45	_{-	0.11	}^{+	0.19	}$&$	-14.79	_{-	0.31	}^{+	0.72	}$&$	-14.34	$&$	-12.63	$&$	-13.57	$&$	-11.98	$&$	0.03	$&$	18.4			$&	3,6	\\
Ursa Major I	&$	97	$&$(	159	,	54	)$&$	4.28	_{-	0.13	}^{+	0.13	}$&$	-13.97	_{-	0.31	}^{+	0.70	}$&$	-13.00	$&$	-12.21	$&$	-	$&$	-12.39	$&$	0.2	$&$	18.3	\pm	0.24	$&	2,5	\\
Carina	&$	105	$&$(	260	,	-22	)$&$	5.63	_{-	0.09	}^{+	0.11	}$&$	-12.69	_{-	0.30	}^{+	0.70	}$&$	-12.07	$&$	-11.53	$&$	-12.47	$&$	-11.72	$&$	4.8	$&$	18.1	\pm	0.23	$&	1,5	\\
Hercules	&$	132	$&$(	29	,	37	)$&$	4.57	_{-	0.14	}^{+	0.14	}$&$	-13.95	_{-	0.32	}^{+	0.71	}$&$	-12.98	$&$	-12.31	$&$	-12.47	$&$	-11.61	$&$	0.4	$&$	18.1	\pm	0.25	$&	2,5	\\
Fornax\tablenotemark{$\dagger$*}	&$	147	$&$(	237	,	-66	)$&$	7.39	_{-	0.14	}^{+	0.14	}$&$	-11.22	_{-	0.32	}^{+	0.70	}$&$	-11.04	$&$	-10.74	$&$	-	$&$	-12.13	$&$	278.7	$&$	18.2	\pm	0.21	$&	2,5	\\
Leo IV	&$	154	$&$(	265	,	57	)$&$	3.93	_{-	0.15	}^{+	0.15	}$&$	-14.72	_{-	0.32	}^{+	0.71	}$&$	-13.88	$&$	-12.79	$&$	-	$&$	-11.97	$&$	0.1	$&$	17.9	\pm	0.28	$&	2,5	\\
Canes Ven. II	&$	160	$&$(	114	,	83	)$&$	3.90	_{-	0.20	}^{+	0.20	}$&$	-14.79	_{-	0.35	}^{+	0.72	}$&$	-13.94	$&$	-12.83	$&$	-	$&$	-12.26	$&$	0.09	$&$	17.9	\pm	0.25	$&	2,5	\\
Columba I	&$	182	$&$(	232	,	-29	)$&$	3.79	_{-	0.07	}^{+	0.13	}$&$	-15.01	_{-	0.29	}^{+	0.70	}$&$	-14.24	$&$	-13.04	$&$	-13.09	$&$	-11.78	$&$	0.07	$&$	17.6			$&	4	\\
Indus II	&$	214	$&$(	354	,	-37	)$&$	3.69	_{-	0.14	}^{+	0.16	}$&$	-15.25	_{-	0.32	}^{+	0.71	}$&$	-14.55	$&$	-13.21	$&$	-12.37	$&$	-11.57	$&$	0.06	$&$	17.4			$&	4	\\
Canes Ven. I	&$	218	$&$(	74	,	80	)$&$	5.48	_{-	0.09	}^{+	0.09	}$&$	-13.48	_{-	0.30	}^{+	0.70	}$&$	-12.81	$&$	-12.24	$&$	-	$&$	-11.98	$&$	3.4	$&$	17.7	\pm	0.26	$&	2,5	\\
Leo II	&$	233	$&$(	220	,	67	)$&$	6.07	_{-	0.13	}^{+	0.13	}$&$	-12.94	_{-	0.31	}^{+	0.70	}$&$	-12.46	$&$	-11.95	$&$	-	$&$	-12.37	$&$	13.3	$&$	17.6	\pm	0.18	$&	2,5	\\
Leo I\tablenotemark{$\dagger$}	 &$	254	$&$(	226	,	49	)$&$	6.69	_{-	0.13	}^{+	0.13	}$&$	-12.40	_{-	0.31	}^{+	0.70	}$&$	-12.08	$&$	-11.68	$&$	-	$&$	-12.20	$&$	55.6	$&$	17.7	\pm	0.18	$&	2,5	\\
Eridanus II\tablenotemark{*}	&$	330	$&$(	250	,	-52	)$&$	4.92	_{-	0.07	}^{+	0.09	}$&$	-14.40	_{-	0.29	}^{+	0.70	}$&$	-13.54	$&$	-12.91	$&$	-14.96	$&$	-11.86	$&$	0.9	$&$	17.3			$&	3,6	\\[1ex]
\tableline\tableline
\end{tabular}
\tablerefs{(1) \citet{wol10}, (2) \citet{kir13}, (3) \citet{bec15}, (4) \citet{drl15_2}, (5) \citet{ack15}, (6) \citet{drl15}}
\end{table*}

\end{document}